\documentstyle[12pt]{article}
\catcode`\@=11
\def\marginnote#1{}
\newcount\hour
\newcount\minute
\newtoks\amorpm
\hour=\time\divide\hour by60
\minute=\time{\multiply\hour by60 \global\advance\minute by-\hour}
\edef\standardtime{{\ifnum\hour<12 \global\amorpm={am}%
        \else\global\amorpm={pm}\advance\hour by-12 \fi
        \ifnum\hour=0 \hour=12 \fi
        \number\hour:\ifnum\minute<10 0\fi\number\minute\the\amorpm}}
\edef\militarytime{\number\hour:\ifnum\minute<10 0\fi\number\minute}
\def\draftlabel#1{{\@bsphack\if@filesw {\let\thepage\relax
   \xdef\@gtempa{\write\@auxout{\string
      \newlabel{#1}{{\@currentlabel}{\thepage}}}}}\@gtempa
   \if@nobreak \ifvmode\nobreak\fi\fi\fi\@esphack}
        \gdef\@eqnlabel{#1}}
\def\@eqnlabel{}
\def\@vacuum{}
\def\draftmarginnote#1{\marginpar{\raggedright\scriptsize\tt#1}}
\def\draft{\oddsidemargin -.5truein
        \def\@oddfoot{\sl preliminary draft \hfil
        \rm\thepage\hfil\sl\today\quad\militarytime}
        \let\@evenfoot\@oddfoot \overfullrule 3pt
        \let\label=\draftlabel
        \let\marginnote=\draftmarginnote
   \def\@eqnnum{(\theequation)\rlap{\kern\marginparsep\tt\@eqnlabel}%
\global\let\@eqnlabel\@vacuum}  }

\def\preprint{\twocolumn\sloppy\flushbottom\parindent 1em
        \leftmargini 2em\leftmarginv .5em\leftmarginvi .5em
        \oddsidemargin -.5in    \evensidemargin -.5in
        \columnsep 15mm \footheight 0pt
        \textwidth 250mmin      \topmargin  -.4in
        \headheight 12pt \topskip .4in
        \textheight 175mm
        \footskip 0pt
        \def\@oddhead{\thepage\hfil\addtocounter{page}{1}\thepage}
        \let\@evenhead\@oddhead \def\@oddfoot{} \def\@evenfoot{} }

\def\titlepage{\@restonecolfalse\if@twocolumn\@restonecoltrue\onecolumn
     \else \newpage \fi \thispagestyle{empty}\c@page\z@
        \def\thefootnote{\fnsymbol{footnote}} }

\def\endtitlepage{\if@restonecol\twocolumn \else  \fi
        \def\thefootnote{\arabic{footnote}}
        \setcounter{footnote}{0}}  

\catcode`@=12
\relax
\def\beq{\begin{equation}}
\def\eeq{\end{equation}}
\def\bea{\begin{array}}
\def\eea{\end{array}}
\def\ov{\overline}
\def\dalpha{{\dot\alpha}}

\def\Re{{\rm Re}\,}
\def\Im{{\rm Im}\,}
\def\crbig{\\\noalign{\vspace {3mm}}}
\def\bigint{{\displaystyle\int}}
\relax

\begin{document}
\topmargin-2.4cm
\begin{titlepage}
\begin{flushright}
NEIP--94--NNN \\
hep--th/9412086 \\
December 1994
\end{flushright}
\vskip 0.25in
\begin{center}{\Large\bf
The Linear Multiplet and Quantum String Effective Actions }
\vskip .2in
{\bf J.-P. Derendinger}
\vskip .1in
Institut de Physique \\
Universit\'e de Neuch\^atel \\
CH--2000 Neuch\^atel, Switzerland
\end{center}
\vskip .2in
\begin{center}
{\bf Abstract}
\end{center}
\begin{quote}
Quantum symmetries of four-dimensional superstrings are frequently
realized in an anomaly-cancellation mode in the effective low-energy
supergravity. The massless antisymmetric tensor plays an important
r\^ole in this mechanism and the choice of its supersymmetric description,
using either a chiral or a linear multiplet,
appears to introduce significant conceptual and practical differences
at the string loop level. This paper reviews the construction
of loop-corrected string effective supergravities with the dilaton and
antisymmetric tensor embedded in a linear multiplet. Using anomaly
cancellation and the linear multiplet allows to obtain an
all-order renormalization-group invariant effective lagrangian
for a pure gauge sector with field-dependent gauge coupling constant.
\end{quote}
\vspace{1.3cm}
\begin{center}
Presented at the Workshop "Physics from Planck Scale to Electroweak
Scale", Warsaw, Poland, September 1994.
\end{center}
\end{titlepage}
\setcounter{footnote}{0}
\setcounter{page}{0}
\newpage
\section{The linear multiplet}

A massless antisymmetric tensor $b_{\mu\nu}$ plays an important r\^ole in
the low-energy effective supergravity of superstrings. Its existence is a
universal property: the gravity sector of four-dimensional superstrings
contains, together with the graviton, $b_{\mu\nu}$ and a real scalar, the
dilaton. It is then natural to take advantage of the particular
properties of the supermultiplet generalizing the antisymmetric tensor
when discussing the structure of the effective supergravity theory.

This section will review some aspects of the supersymmetric description
of the antisymmetric tensor. For simplicity, the discussion will first
concentrate on global supersymmetry. The case of supergravity which will be
of importance for the superstring case will be introduced at the end
of the section.

In global supersymmetry, the description of the antisymmetric tensor uses
two superfields\cite{linear},
the prepotential $\psi^\alpha$ and the linear multiplet
$L$. The prepotential is a chiral, spinor (the index $\alpha$) superfield
with the gauge transformation
\beq
\label{pregauge}
\delta\psi^\alpha = i\ov{\cal DD}{\cal D}^\alpha K,
\eeq
where $K$ is a real vector superfield. This transformation can be used
to choose a gauge in which some of the components of $\psi^\alpha$
are eliminated. In this gauge, $\psi^\alpha$ contains the antisymmetric
tensor, a real scalar $C$ and a spinor $\chi$. And (\ref{pregauge}) reduces
to the residual (bosonic) gauge symmetry
\beq
\label{bsym}
\delta b_{\mu\nu} = \partial_\mu k_\nu - \partial_\nu k_\mu,
\eeq
which preserves the gauge choice.

The real linear multiplet\cite{linear} is the supersymmetrization of the
field strength
\beq
\label{vmu}
v_\mu = {1\over\sqrt2} \epsilon_{\mu\nu\rho\sigma}\partial^\nu b^{\rho\sigma}
\eeq
of $b_{\mu\nu}$. In terms of the prepotential, it is defined by
\beq
\label{Ldef1}
L = {\cal D}^\alpha \psi_\alpha + \ov{\cal D}_\dalpha \ov\psi^\dalpha,
\eeq
which is invariant under the transformation (\ref{pregauge}).
Alternatively, it can be obtained by imposing the constraints
\beq
\label{Ldef2}
{\cal DD} L = \ov{\cal DD} L = 0,
\eeq
on a real vector superfield. The component expansion of $L$ is
$$
L = C + i\theta\chi -i\ov\theta\ov\chi +\theta\sigma^\mu\ov\theta v_\mu
-{1\over2}\theta\theta\ov\theta(\partial_\mu\chi\sigma^\mu)
-{1\over2}\ov{\theta\theta}\theta(\sigma^\mu\partial_\mu\ov\chi)
-{1\over4}\theta\theta\ov{\theta\theta}\Box C.
$$
It is interesting to remark that $L$ does not contain any auxiliary field.
A lagrangian involving the linear multiplet $L$ only
does not include a scalar potential and supersymmetry cannot spontaneously
break.

A generic lagrangian for $L$ and chiral matter denoted collectively by
$\Sigma$ is of the form
$$
{\cal L}_{global} = 2\int d^2\theta d^2\ov\theta\, \Phi(L,\Sigma, \ov\Sigma)
+ \int d^2\theta\, w(\Sigma) + \int d^2\ov\theta\, \ov w(\ov\Sigma),
$$
with an arbitrary real function $\Phi$ and a superpotential $w$ independent
from $L$. Coupling this theory to gauge fields leads to
\beq
\label{Llag2}
{\cal L}_{global} = 2\int d^2\theta d^2\ov\theta\, \Phi(\hat L,
\Sigma, \ov\Sigma e^V)
+ \int d^2\theta\, w(\Sigma) + \int d^2\ov\theta\, \ov w(\ov\Sigma),
\eeq
where $V = V^aT^a$ is the Lie-algebra-valued
vector superfield of gauge potentials,
the superpotential is a gauge-invariant analytic function,
\beq
\label{hatL}
\hat L = L - 2\Omega,
\eeq
and $\Omega$ is the Chern-Simons superfield. This last object can be defined
by its relation with the chiral superfield of gauge curvatures,
$$
W^\alpha = -{1\over4}\ov{\cal DD}e^{-V}{\cal D}^\alpha e^V = W^{\alpha\,a}T^a,
$$
which reads
\beq
\label{CSdef}
\ov{\cal DD}\Omega = W^aW^a, \qquad {\cal DD}\Omega= \ov W^a \ov W^a.
\eeq

Since theory (\ref{Llag2}) only depends on $b_{\mu\nu}$ through its curl
$v_\mu$ [eq. (\ref{vmu})], a duality transformation can be performed to
replace $b_{\mu\nu}$ by a (pseudo)scalar field. With supersymmetry, the
duality transformation replaces $L$ by a chiral superfield, an operation
which will be discussed in detail in the next section.

For reasons to be explained below, conformal
supergravity\cite{conf} offers the
appropriate formalism to extend the global lagrangian (\ref{Llag2}) to
local supersymmetry. One needs to introduce a chiral
compensating supermultiplet $S_0$ with conformal and chiral
weights equal to one\footnote{This choice of compensator leads to the
old minimal set of supergravity auxiliary fields,
which allows the most general matter
couplings\cite{FGKVP}.}.
Since the conformal weight of $L$ is two, the supergravity lagrangian
generalizing (\ref{Llag2}) is
\beq
\label{sugra1}
{\cal L}_{local} = \left[ S_0\ov S_0 \Phi({\hat L\over S_0\ov S_0},
\Sigma, \ov\Sigma e^V)\right]_D + \left[S_0^3w(\Sigma)\right]_F.
\eeq
The real and chiral density formula are denoted respectively by
$[\ldots]_D$ and $[\ldots]_F$. The Poincar\'e supergravity
theory is obtained by imposing in (\ref{sugra1})
gauge-fixing conditions for the superconformal symmetries not contained
in the super-Poincar\'e algebra. These gauge choices use in particular
the scalar and fermion components $z_0$ and $\psi_0$ of
$S_0$, and leave the auxiliary
component $f_0$ and the gauge field $A_\mu$ of the chiral, internal $U(1)$
of the superconformal algebra as supergravity auxiliary fields.

It is worth recalling that the linear multiplet has made at least two
appearances in past developments of supergravity theories. Firstly, `new
minimal supergravity' is obtained in the conformal framework using a linear
compensating multiplet\cite{dWR,FGKVP}, with lagrangian $[L\log L]_D$. And
secondly, early investigations\cite{LLO} of the tree-level effective
supergravity of superstrings suggested the relevance of `16+16
supergravity'\cite{1616}, which corresponds in the conformal approach
to a theory with a linear and a chiral
multiplet, with some particular gauge fixing\cite{16162}.

\section{Gauge kinetic lagrangian and gauge coupling constants}

Since field-dependent gauge coupling constants in superstring effective
supergravities will be at the center of the discussion, we briefly
consider\cite{DQQ}
the sector of gauge kinetic terms in the global lagrangian (\ref{Llag2})
and in its supergravity extension (\ref{sugra1}).

In global supersymmetry, the expansion of the Chern-Simons superfield
in the Wess-Zumino gauge contains the following terms:
$$
\begin{array}{rcl}
\Omega &=& -{1\over2}\theta\theta\ov{\theta\theta}\left( {1\over2}D^aD^a
-{1\over2}i\lambda^a\sigma^\mu\partial_\mu\ov\lambda^a
+{1\over2}i\partial_\mu\lambda^a\sigma^\mu\ov\lambda^a
-{1\over4} F^{a\,\mu\nu}F^a_{\mu\nu}\right) \crbig
&&+ {1\over2}(\theta\sigma^\mu\ov\theta) \epsilon^{\mu\nu\rho\sigma}
\omega_{\nu\rho\sigma} + {1\over4}\theta\theta\ov\lambda^a\ov\lambda^a
+{1\over4}\ov{\theta\theta}\lambda^a\lambda^a + \ldots,
\end{array}
$$
where $\omega_{\nu\rho\sigma}$ is the gauge Chern-Simons form, normalised by
$$
\partial_\mu\epsilon^{\mu\nu\rho\sigma}\omega_{\nu\rho\sigma}
={1\over4}\epsilon^{\mu\nu\rho\sigma}F^a_{\mu\nu}F^a_{\rho\sigma}
={1\over2}\tilde F^a_{\mu\nu} F^{a\,\mu\nu}.
$$
Its highest component is the super-Yang-Mills lagrangian (multiplied
by $-1/2$) and it does not contain any $\tilde F^a_{\mu\nu}
F^{a\,\mu\nu}$ contribution. Gauge kinetic terms in lagrangian (\ref{Llag2})
will then clearly be of the form
$$
2\Phi_C \left( {1\over2}D^aD^a
-{1\over2}i\lambda^a\sigma^\mu\partial_\mu\ov\lambda^a
+{1\over2}i\partial_\mu\lambda^a\sigma^\mu\ov\lambda^a
-{1\over4} F^{a\,\mu\nu}F^a_{\mu\nu}\right),
$$
with $\Phi_C= {\partial\over\partial C}\Phi(C,z,\ov z)
= \left[{\partial\over\partial L}\Phi(\hat L,\Sigma,\ov\Sigma e^V)\right]
_{\theta=\ov\theta=0}$. The field-dependent gauge coupling
constant is then\cite{DFKZ1,DQQ}
\beq
\label{g1}
{1\over g^2} = 2\Phi_C.
\eeq
With the linear multiplet, the gauge coupling constant is a real
function which is not constrained  to be harmonic. It is obtained
in a real $D$-term, contrary to the chiral case where it is related to
a holomorphic function. This illustrates the fact that holomorphicity
properties of gauge couplings do not follow from supersymmetry only but
require also a selection of supersymmetric multiplets\cite{DFKZ1}.

Lagrangian (\ref{Llag2}) also contains a term
\beq
\label{CSterm}
-\left[\Phi_{CC}v_\mu -i\Phi_{Cz}\partial_\mu z +i\Phi_{C\ov z}
\partial_\mu\ov z\right]\epsilon^{\mu\nu\rho\sigma}\omega_
{\nu\rho\sigma}.
\eeq
This contribution will only give rise, by partial integration, to an
$\tilde F^a_{\mu\nu}F^{a\,\mu\nu}$ with the very special choice
\beq
\label{case}
\Phi= \hat L[\phi(\Sigma) + \ov\phi(\ov\Sigma) ].
\eeq
In this case, $\Phi_{CC}= 0$, $\Phi_{Cz}\partial_\mu z=
\partial_\mu\phi(z)$, $\Phi_{C\ov z}\partial_\mu\ov z =
\partial_\mu \ov\phi(\ov z)$, and (\ref{CSterm}) becomes
$$
\begin{array}{rcl}
i\partial_\mu[\phi(z) - \ov\phi(\ov z)] \epsilon^{\mu\nu\rho\sigma}
\omega_{\nu\rho\sigma}
&=& 2\Im\phi(z) \partial_\mu\epsilon^{\mu\nu\rho\sigma}\omega_
{\nu\rho\sigma} + {\rm total\,\,derivative} \crbig
&=& \Im\phi(z) \tilde F^a_{\mu\nu}F^{a\,\mu\nu}
+ {\rm total\,\,derivative}.
\end{array}
$$
In this case, the gauge coupling constant is proportional to
the real part of $\phi(z)$:
$$
{1\over g^2} = 4\Re\phi(z).
$$
Notice however that since
$\int d^2\theta d^2\ov\theta\, L[\phi(\Sigma)+\ov\phi(\ov\Sigma)] =
$ total derivative,
the choice (\ref{case}) completely eliminates the linear multiplet.
Then,
$$
{\cal L} = -4\int d^2\theta d^2\ov\theta\,
\Omega[\phi(\Sigma)+\ov\phi(\ov\Sigma)]
= \int d^2\theta\, \phi(\Sigma)W^a W^a + {\rm h.c.},
$$
which is the usual super-Yang-Mills theory with
gauge coupling depending on chiral matter. The fact that
the gauge coupling is the real part of a holomorphic function appears then
to depend on the description of matter with chiral multiplets only.

\section{Supersymmetric duality}

Any theory with a linear multiplet can be in
principle transformed into a theory
with chiral multiplets only, using a supersymmetric duality transformation.
Suppose we want to transform (\ref{Llag2}) into a theory with a chiral
superfield $S$ replacing $L$. The starting point is to write
a classically equivalent theory, of the form:
\beq
\label{dual1}
{\cal L}_1 = 2\int d^2\theta d^2\ov\theta\, \Phi(\hat L-U,\Sigma,\ov
\Sigma e^V)  + \left\{\int d^2\theta\, \left[ -{1\over2}S\ov{\cal DD}U +
w(\Sigma)\right] +{\rm h.c.} \right\}.
\eeq
This lagrangian possesses a new gauge invariance under
\beq
\label{newgauge}
U \longrightarrow U + {\cal U},  \qquad\qquad
L \longrightarrow L + {\cal U},
\eeq
where the gauge parameter is a linear superfield ${\cal U}$: ${\cal DD}
{\cal U}=\ov{\cal DD}{\cal U}=0$. The equation of motion for $S$ implies that
$U$ is a linear multiplet and (\ref{Llag2}) is then simply the gauge choice
$U=0$ in (\ref{dual1}), after solving for $S$. Notice that a superfield $U$
submitted to the gauge transformation (\ref{newgauge}) can be regarded
as the supersymmetric extension of an antisymmetric tensor $a_{\mu\nu\rho}$
with gauge variation $\delta a_{\mu\nu\rho}=
\partial_{[\mu}{\cal U}_{\nu\rho]}$. The expansion of the superfield
$L-U$ contains
$$
{1\over\sqrt2}(\theta\sigma^\mu\ov\theta) \epsilon_{\mu\nu\rho\sigma}
D^{[\nu} b^{\rho\sigma]},
$$
with a derivative
$$
D^{[\nu} b^{\rho\sigma]} = \partial^{[\nu} b^{\rho\sigma]}
- a^{\nu\rho\sigma}
$$
which is invariant under $\delta b_{\mu\nu} = {\cal U}_{\mu\nu}$.
Theory (\ref{dual1}) can be rewritten
\beq
\label{steps}
\begin{array}{rcl}
{\cal L}_2 &=&
2\bigint d^2\theta d^2\ov\theta\, \left[\Phi(\hat L-U,\Sigma,\ov
\Sigma e^V) - (S+\ov S)(\hat L-U) -2(S+\ov S)\Omega \right]
\crbig
&& +\left\{ \bigint d^2\theta\, w(\Sigma)
+{\rm h.c.} \right\}
\crbig
&=&
2\bigint d^2\theta d^2\ov\theta\, \left[\Phi(\hat L-U,\Sigma,\ov
\Sigma e^V) - (S+\ov S)(\hat L-U) \right]
\crbig
&& +\left\{ \bigint d^2\theta\, \left[ SW^aW^a + w(\Sigma)\right]
+{\rm h.c.} \right\}.
\end{array}
\eeq
The equation of motion for $U$ (or $\hat L-U$) can in principle be used
to express $\hat L-U$ as a function of $S+\ov S$, to finally obtain the
dual theory
\beq
\label{Stheory}
{\cal L}_S = \int d^2\theta d^2\ov\theta\, K(S+\ov S, \Sigma, \ov\Sigma
e^V) + \left\{ \bigint d^2\theta\, \left[ SW^aW^a + w(\Sigma)\right]
+{\rm h.c.} \right\}.
\eeq
In this expression, which possesses the symmetry
$S \longrightarrow S+i$(real constant), the gauge coupling constant is always
\beq
\label{gdual}
{1\over g^2} = 4\Re s,
\eeq
$s$ being the scalar lowest component of $S$. And there is a term of the
form
\beq
\label{SFFdual}
\Im s \tilde F^a_{\mu\nu} F^{a\,\mu\nu} =
-2(\partial_\mu\Im s)\epsilon^{\mu\nu\rho\sigma}\omega_{\nu\rho\sigma}
+{\rm total\,\,derivative}.
\eeq
The duality transformation is performed by solving the equations of motion
for the components $(c,\tilde\chi,m,n,\tilde v_\mu,
\tilde\lambda,d)$ of $U$, for instance in the gauge $L=0$.
In particular, these equations indicate that\cite{DQQ}
\beq
\begin{array}{rcl}
\Re s &=& {1\over2}\Phi_c  , \crbig
\partial_\mu \Im s &=& {1\over 2} \Phi_{cc}\tilde v_\mu -{i\over2}\Phi_{cz}
\partial_\mu z +{i\over2}\Phi_{c\ov z}\partial_\mu \ov z
+ {\rm fermionic\,\,terms}.
\end{array}
\eeq
These relations show the equivalences of (\ref{gdual}) and (\ref{SFFdual})
with respectively (\ref{g1}) and (\ref{CSterm}).

These results have a straightforward extension to conformal
and Poincar\'e super\-gra\-vities\cite{DQQ}. Equation (\ref{gdual}),
in particular, remains true in the local case.
It provides an important information on the description
of superstring effective supergravities. The two dual theories have a different
physical interpretation. Working with the chiral theory, with $S$, implies
that the gauge coupling constant appearing in the lagrangian is the
expectation value of $\Re s$. This field is then directly related to a
bare, unphysical quantity. The physical (renormalized) gauge coupling
constant, which appears in the effective action (the generating functional
of one-particle irreducible Green's functions), is a function of
$\Re s$ (as well as other fields, in general). It is certainly not given
by the expectation value of $\Re s$.
Then, since $\Re s$ is related to a bare quantity of the effective
field theory, it is not necessarily in simple and natural relation with
string parameters at the (string) quantum level.
And there is no reason to expect that quantum symmetries of
the superstring have a natural action on this field.

\section{Superstring effective lagrangians}

The effective supergravity of superstrings describes the
low-energy\footnote{With
respect to the inverse string length and compactification radius.} physics
of string massless modes. It then contains an ultraviolet physical cutoff,
above which `string physics has been integrated'. The effective supergravity
can be defined using a Wilson lagrangian ${\cal L}_W$.
At the level of superstring perturbation theory,
in which supersymmetry does not break, the Wilson effective lagrangian has
a string loop expansion
\beq
\label{LWexp}
{\cal L}_W = \sum_{N\ge0}{\cal L}_W^{(N)}.
\eeq
The Wilson lagrangian and the corresponding action
$S_W=\int d^4x\,{\cal L}_W$ should not be confused with
the more familiar effective action of quantum field theory,
$$
S_\Gamma = \int d^4x\, {\cal L}_\Gamma,
$$
which is the generating functional of one-particle irreducible Green's
functions. While ${\cal L}_W$ is a quantized lagrangian, ${\cal L}_\Gamma$
is a functional of classical fields.

For large classes of superstrings, the tree-level lagrangian
${\cal L}_W^{(0)}$ is well understood. Since the present
discussion focusses on the r\^ole of the antisymmetric
tensor described by a linear multiplet $L$, or on the dual theory with a
chiral $S$, it will be sufficient to use
\beq
\label{LWtree}
{\cal L}_W^{(0)} = -{1\over\sqrt2} \left[ \hat L\left( {\hat Le^{\hat K/3}
\over S_0\ov S_0}\right)^{-3/2}\right]_D +[S_0^3 w]_F,
\eeq
as tree-level Wilson lagrangian\cite{CFV}.
The K\"ahler function $\hat K$ and the superpotential $w$ depend
in a gauge invariant way on the chiral multiplets describing moduli
and charged matter. Using (\ref{Stheory}), theory (\ref{LWtree})
is, by duality, equivalent to the chiral lagrangian
\beq
\label{LWStree}
{\cal L}_S^{(0)} =
-{3\over2}\left[ S_0\ov S_0 e^{-{\cal K}/3} \right]_D
+[SW^aW^a + S^3_0 w]_F,
\qquad
{\cal K} = -\log(S+\ov S) +\hat K.
\eeq
The tree-level Wilson lagrangian possesses the (formal) K\"ahler symmetry
\beq
\label{Kahler}
\begin{array}{rclcrcl}
S_0 &\longrightarrow& e^{\varphi/3} S_0, &\qquad&
w &\longrightarrow& e^{-\varphi} w, \crbig
\hat K &\longrightarrow& \hat K + \varphi + \ov\varphi,  &\qquad&
\hat L &\longrightarrow& \hat L,
\end{array}
\eeq
where $\varphi$ is an arbitrary holomorphic
function of chiral multiplets.

The lagrangian (\ref{LWtree}) is superconformal invariant. Two different
choices of compensator fixings are of interest. Imposing canonical
normalisation of the Einstein term leads to
$$
e^{-1}{\cal L}_W^{(0)} = -{1\over2\kappa^2} R -{1\over4}(\kappa^2 C)^{-1}
F^a_{\mu\nu}F^{a\,\mu\nu} +{1\over4}\kappa^2 (\kappa^2 C)^{-2} v_\mu v^\mu
+\ldots,
$$
and the gauge coupling constant is
\beq
\label{gtree}
g_{tree}^{-2} = (\kappa^2 C)^{-1} = 4\Re s.
\eeq
The second choice is the `string frame' where the same bosonic terms become
$$
e^{-1}{\cal L}_W^{(0)} = {\varphi^{-1/2}\over\kappa^2}\left[
-{1\over2} R -{1\over4}M_s^{-2} F^a_{\mu\nu}F^{a\,\mu\nu}
+{1\over4}M_s^{-4} v_\mu v^\mu\right]
+\ldots
$$
where $\varphi=\kappa^2 C$ is the string `loop-counting parameter'
and $M_s^2 = \langle C\rangle$ is the string scale. With this choice,
$$
g_{tree}^{-2} = (\kappa^2 C)^{-1} = (4\Re s)^{2/3},
$$
since the relation between $\Re s$ and $C$ depends on the superconformal
gauge fixing.

\newpage
\section{One loop: a string symmetry, anomaly cancellation}

Some contributions to ${\cal L}_W^{(1)}$, in eq. (\ref{LWexp}), have been
revealed by the effective field-theory description of string one-loop
calculations of gauge kinetic terms in $(2,2)$ symmetric
orbifolds\cite{DKL,ANT,MS}. The moduli dependence of these corrections
is in large part dictated by target-space duality, a perturbative string
symmetry. In the effective supergravity, the crucial
observation\cite{DFKZ1} is that since target-space
duality is an anomalous symmetry of ${\cal L}^{(0)}$, invariance is
restored by an anomaly-cancellation
mechanism due to specific contributions in ${\cal L}^{(1)}$.
This mechanism is similar to the cancellation of gauge and gravitational
anomalies in ten-dimensional superstrings\cite{GS}.
Target-space duality acts on the tree-level lagrangian (\ref{LWtree}) like a
K\"ahler transformation (\ref{Kahler}). The cancellation of target-space
duality anomaly is then a mechanism of K\"ahler anomaly cancellation
in supergravity\cite{CO}.

This mechanism will be reviewed here in its simplest version only.
Specifically, we will consider the case of vanishing
threshold corrections\footnote{
This means that anomaly cancellation also cancels the one-loop
modulus-dependent correction
to physical gauge coupling constants.}. We will also for simplicity
restrict the discussion to a simple gauge group, which could be the
hidden $E_8$ group present in symmetric $(2,2)$ orbifolds. The
discussion of the general case can be found in the original
literature\cite{DFKZ1,DLLI}.

The nature of K\"ahler symmetry is easily elucidated. Clearly, it is a
sigma-model symmetry: it acts on $\hat K$ and $\hat K$ is the K\"ahler
potential of the supersymmetric sigma-model defining the couplings of chiral
multiplets. But in supergravity, it also acts as an R-symmetry which,
in particular, rotates all fermions in the theory. This is due to the presence
of the supergravity auxiliary field $A_\mu$, which in the superconformal theory
is the gauge field of the internal chiral $U(1)$. This chiral group does not
commute with supersymmetry transformations and hence acts like an R-symmetry.
The equation of motion of its gauge field is
$$
A_\mu = {i\over3}\left[{\partial\hat K\over\partial z} (\partial_\mu z)
- {\partial\hat K\over\partial\ov z} (\partial_\mu \ov z)\right]
+\,\,\ldots\,\,,
$$
for an arbitrary chiral scalar component $z$. The K\"ahler transformation
(\ref{Kahler}) is then a chiral $U(1)$ gauge transformation of $A_\mu$, once
the equation of motion for $A_\mu$ has been solved, and $\hat K$ can be
assimilated to the (superfield) K\"ahler connection.

The existence of chiral couplings of the K\"ahler connection to fermions
implies that a non-zero anomalous triangle graph with, for instance, two
external gauge fields and one external K\"ahler connection will in general
arise at one-loop. This diagram corresponds to a non-local contribution to the
effective lagrangian ${\cal L}_\Gamma$ (not to ${\cal L}_W$), which,
in the absence of chiral charged matter as assumed here, is of the form
\beq
\label{anom1}
\Delta_{triangle} =
-{A\over4}\left[ {1\over3}W^aW^a{\cal P}_C \hat K\right]_F,
\eeq
where ${\cal P}_C$ is the non-local chiral projector.
Gauginos only contribute to
the anomaly and its coefficient is $A= 3C(G)/(8\pi^2)$, $C(G)$ being
the quadratic Casimir of the gauge group $G$.
The transformation of $\Delta_{triangle}$ under (\ref{Kahler}) is
\beq
\label{varanom1}
\delta(\Delta_{triangle}) =
-{A\over4}\left[ {1\over3}W^aW^a\varphi \right]_F,
\eeq
a local expression.

The one-loop Wilson lagrangian ${\cal L}_W^{(1)}$ should then contain a
K\"ahler-variant term able to cancel (\ref{varanom1}) by its variation.
This contribution is\cite{DFKZ1}
\beq
\label{GS1}
{\cal L}_W^{(1)} = {A\over3}{1\over4}\left[ \hat L\hat K\right]_D,
\eeq
which includes a coupling of the antisymmetric tensor to the K\"ahler
connection\footnote{The existence of this coupling has been demonstrated
by direct string calculations\cite{ANT}.}.
With this one-loop correction, the Wilson lagrangian becomes
\beq
\label{LWone}
{\cal L}_W^{(0)} + {\cal L}_W^{(1)} =
\left[ -{1\over\sqrt2} \hat L\left( {\hat Le^{\hat K/3}
\over S_0\ov S_0}\right)^{-3/2}
+ {A\over12}\hat L\hat K \right]_D +[S_0^3 w]_F.
\eeq
Repeating the steps (\ref{steps}) and (\ref{Stheory})
to perform the duality transformation leads to the substitution
\beq
\label{subst}
S+\ov S \,\,\longrightarrow\,\, S+\ov S -{A\over3}\hat K
\eeq
in the tree-level K\"ahler function ${\cal K}$ given in eq.
(\ref{LWStree}). This result is reminiscent of the mechanism cancelling
gauge anomalies for $U(1)$ factors of the gauge group, where the substitution
is
$$
S+\ov S \,\,\longrightarrow\,\, S+\ov S -\alpha V,
$$
$V$ being the vector superfield of the $U(1)$ gauge potential and $\alpha$
the coefficient of the anomaly\cite{DSW}.

\section{An effective all-order gauge sector}

In the previous section, the information that target-space duality is
a quantum string symmetry has been used to derive a one-loop contribution
to the Wilson effective lagrangian, using anomaly cancellation as a
basic principle. The one-loop lagrangian (\ref{LWone}) has been
written in the superconformal formalism. It
is then also scale invariant as long as a superPoincar\'e invariant
gauge choice has not been taken. It is tempting to
extend\cite{DFKZ2} the method applied
to restore the anomalous target-space duality also to the dilatation
symmetry contained in the conformal algebra. Since the
gauge beta-function corresponds\cite{anom} to the
dilatation anomaly obtained when
varying the compensator field $S_0$, the superconformal
formalism used here is certainly appropriate.

The discussion given in this section only
applies to a super-Yang-Mills theory without chiral charged matter.
This situation corresponds, for instance, to
the hidden $E_8$ sector of $(2,2)$ superstrings. The reason for this
limitation is not only simplicity. The generalization to a string gauge
sector with chiral charged matter and a non simple gauge group is not
known at present. Furthermore, the all-order
field theory results on beta-functions, which are necessary to check
the results of the
construction, are not available with arbitrary chiral matter.

To proceed, consider the effect of a scale change on
the gauge coupling constant. Under
$$
M \longrightarrow \lambda M,
$$
the variation of the one-loop gauge coupling constant is
$$
\delta(g^{-2}) = A\log\lambda,
$$
where $A=3C(G)/(8\pi^2)$ is the coefficient of the one-loop beta
function which already appeared in the K\"ahler/target-space duality
anomaly (\ref{anom1}). In an effective
lagrangian formalism, this anomalous variation reads
\beq
\label{varia}
\delta\left( -{1\over4}{1\over g^2}F_{\mu\nu}^aF^{a\,\mu\nu}\right) =
-{1\over4}A\log\lambda F^a_{\mu\nu}F^{a\,\mu\nu}.
\eeq
The next step is to construct a supersymmetric expression such that its
variation compensates in particular (\ref{varia}).
This should be done without introducing any new K\"ahler anomaly. Using the
same notation as in the preceding section, there exists\cite{DFKZ2,DQQ} a
unique supersymmetric expression with a variation including (\ref{varia}):
\beq
\label{Lterm}
{A\over4}\left[ \hat L\log(\hat L/\mu^2)\right]_D.
\eeq
Its anomaly behaviour follows from the conformal weight two of $\hat L$.
The quantity $\mu$ is an arbitrary number. It is not
a scale since a constant supermultiplet must have zero conformal dimension.
It is only after conformal gauge fixing that $\mu$ will acquire the r\^ole of
a reference scale, as will be seen below.

Collecting its tree, one-loop and running contributions,
the complete Wilson lagrangian is
\beq
\label{Lall}
{\cal L}_W = -{1\over\sqrt2}\left[ \hat L\left({\hat Le^{\hat K/3}\over S_0
\ov S_0}\right)^{-3/2}+
{A\over4}\hat L\log\left({\hat Le^{\hat K/3}\over \mu^2}\right)
\right]_D + \left[S_0^3w\right]_F.
\eeq
Notice that the anomaly term can also be written:
$$
{A\over4}\left[ \hat L\log\left({\hat Le^{\hat K/3}\over S_0\ov S_0}\right)
+\hat L\log\left({S_0\ov S_0\over\mu^2} \right)\right]_D.
$$
This last expression shows that both K\"ahler and dilatation anomalies
are conveniently carried by the chiral compensator $S_0$, a natural
observation
since $S_0$ has chiral {\it and} Weyl weights equal to one and K\"ahler
symmetry finds its superconformal origin in the chiral internal $U(1)$.

The effective lagrangian ${\cal L}_\Gamma$ can be easily obtained
from the
Wilson theory (\ref{Lall}). It differs from ${\cal L}_W$ (viewed
as a functional of classical fields instead of an operator-valued lagrangian)
by contributions arising from loop diagrams, which are partially non-local:
\beq
\label{Leff}
{\cal L}_\Gamma =
{\cal L}_W - {A\over4}\left[ W^aW^a{\cal P}_C\log \left({\hat Le^{\hat K/3}
\over \mu^2}\right) \right]_F +\ldots,
\eeq
the dots indicating contributions unrelated to gauge kinetic terms which are
not of interest here.
As it should, it is invariant under both K\"ahler and scale (conformal)
transformations.

The Poincar\'e invariant theory is obtained by a superconformal
gauge fixing which uses in particular the scalar $z_0$ and fermionic $\psi_0$
components of the compensating multiplet $S_0$. The gravitational lagrangian
will take the canonical form $-{1\over2\kappa^2}eR$ if one chooses
\beq
\label{Udef}
U \equiv {1\over\sqrt2}\left({Ce^{\hat K/3}\over z_0\ov z_0}\right)^{-3/2}
 ={1\over\kappa^2 C} - {A\over6},
\eeq
which defines $|z_0|$ and introduces the Planck scale via $\kappa =
\sqrt {8\pi G_N} = (M_P/8\pi)^{-1}$. Notice that the quantity $U$ is K\"ahler
and scale invariant.

The field-dependent Wilson gauge coupling constant is defined as the
coefficient of gauge kinetic terms in ${\cal L}_W$:
\beq
\label{gWilson}
{\cal L}_W = -{1\over4}\,{1\over g^2_W}\, F^a_{\mu\nu}F^{a\,\mu\nu} + \ldots
\eeq
It is a bare, unphysical quantity, like all parameters appearing in
${\cal L}_W$. In theory (\ref{Lall}), one finds,
\beq
\label{gWilson2}
{1\over g_W^2(z_0\ov z_0)} = U -{A\over3}\log U
+{A\over2}\log\left({z_o\ov z_0\over \mu^2}\right) + c,
\eeq
where the constant $c={A\over2}(1-{1\over3}\log2)$ can be
absorbed in $\mu$.
The Wilson coupling is neither K\"ahler nor scale invariant.
Notice also that the Poincar\'e
gauge fixing promotes the quantity $\mu$ to a scale, once
$|z_0|$ is measured in units of the Planck scale.

The field-dependent effective gauge coupling constant is defined as the
coefficient of gauge kinetic terms in the effective lagrangian:
\beq
\label{geff}
{\cal L}_\Gamma = -{1\over4}{1\over g^2_\Gamma} F^a_{\mu\nu}F^{a\,\mu\nu}
+ \ldots
\eeq
The non-local contributions to ${\cal L}_\Gamma$ in eq. (\ref{Leff})
include local gauge kinetic terms, and one simply obtains
\beq
\label{geff2}
{1\over g^2_\Gamma(\mu)} = U,
\eeq
a K\"ahler and target-space duality invariant result.
According to (\ref{Udef}), the physical effective
gauge coupling constant in the loop-corrected theory is specified by
the expectation value of the real, dimensionless field
$\kappa^2 C$\footnote{The field
$\kappa^2 C$ is the lowest component of $\kappa^2\hat L$ in the Wess-Zumino
gauge.}. And, as discussed in section 2, the Wilson lagrangian
(\ref{Lall}) does not contain any $F_{\mu\nu}^a \tilde F^{a\,\mu\nu}$ term.

The dependence on the arguments $z_0\ov z_0$ and $\mu$ of $g_W$
and $g_\Gamma$ follows from the following remark.
Using eqs. (\ref{gWilson2}) and (\ref{geff2}), one finds:
\begin{eqnarray}
\label{rggamma}
\displaystyle{\mu{\partial\over\partial\mu} g_\Gamma} &=&
\displaystyle{ -{3C(G)\over16\pi^2}
g_\Gamma^3 \,\left[1-{C(G)\over8\pi^2}g_\Gamma^2\right]^{-1},} \crbig
\label{rgwilson}
\displaystyle{|z_0|{\partial\over\partial |z_0|} g_W}&=&
\displaystyle{-{3C(G)\over16\pi^2}g_W^3}.
\end{eqnarray}
The scale dependence of the physical coupling $g_\Gamma$ coincides
with the all-order renor\-ma\-li\-sation-group
equation in pure super-Yang-Mills
theory\cite{allorder}. And the Wilson gauge coupling is only affected by
one-loop running. Relation (\ref{rgwilson}) leads to
\beq
\label{gWilsonmu2}
{1\over g_W^2(\mu^2)} = {1\over g^2_\Gamma(\mu)} +
{A\over3} \log g_\Gamma^2 (\mu).
\eeq
Equations (\ref{rgwilson}) and (\ref{gWilsonmu2}) have been established
in pure super-Yang-Mills theories by Shifman and Vainshtein \cite{SV}. They
have been obtained here from anomaly cancellation in an effective lagrangian
construction.

Finally, the dual Wilson lagrangian with $L$ replaced by the
chiral multiplet $S$ has a gauge coupling constant given by
\beq
\label{gdual2}
{1\over g_W^2(z_0\ov z_0)} = 4\Re s,
\eeq
as in (\ref{gdual}).
The duality transformation is performed by expressing $U$ as a function of
$\Re s$, using eqs. (\ref{gWilson2}) and (\ref{gdual2}). The solution can
however only be found in a perturbative expansion: while the linear multiplet
theory (\ref{Lall}) is known exactly, with all-order gauge couplings
(\ref{gWilson2}) and (\ref{geff2}), the dual chiral theory can
only be expressed as an infinite series.

\section{Conclusions}

Formally, a supergravity theory with a linear multiplet can
always be transformed into a theory with chiral multiplets only,
using duality. String effective
supergravities indicate however that this formal equivalence does not
imply conceptual equivalence, besides the fact that duality
can be hard or impossible to perform analytically.
The duality transformation is applied to the Wilson lagrangian,
with the loop expansion (\ref{LWexp}). Since the scalar component $C$ of the
linear multiplet appears to be the string loop-counting parameter, it
seems natural to formulate the loop expansion with $L$.
The relation between $S$ and $L$
changes order by order. Duality will then induce a rearrangement
of perturbation theory which can obscure the identification of string
parameters and amplitudes with the
fields of the effective supergravity. This is exemplified by the
effective description of thereshold corrections
in orbifolds\cite{DKL,DFKZ1}, and also by substitution (\ref{subst}),
at one loop.

The construction of the effective gauge sector provides
a number of interesting results. The all-order renormalization-group
behaviour is obtained from anomaly cancellation. In some sense,
there is a non-renormalization property similar to an Adler-Bardeen theorem.
The field-dependent gauge coupling, a physical parameter,
is the expectation value of the real field
$C$. The chiral version of the theory uses a complex scalar $s$, which
gives the bare unphysical Wilson coupling. And the linear theory does not
include any $F\tilde F$ term.

Finally, the strength of the gauge interaction
and the scale at which this force
becomes confining are characterized by the renormalization-group invariant
scale $\Lambda$, given by the real superfield
$$
\Lambda^3 = \mu^3 Ue^{-3U/A},
$$
which is also target-space duality invariant. This is also the scale at
which dynamical supersymmetry breaking is expected to be induced by
gaugino condensates. The gauge sector defined by the Wilson
lagrangian (\ref{Lall}) can be used for an effective lagrangian treatment of
this phenomenon\cite{BDQQ}.

\end{document}